\documentclass[10pt,twocolumn, notitlepage]{article} 
\usepackage{graphicx} 
\usepackage{subcaption}
\usepackage{amsfonts}
\usepackage[T1]{fontenc}
\usepackage{booktabs}

\usepackage{hyperref}

\AtBeginDocument{%
  }

\usepackage{geometry}
\begin{document}

\title{An Analysis of the Impact of Gold Open Access Publications \\ in Computer Science}

\author{P\'{a}draig Cunningham, Barry Smyth\thanks{Both authors contributed equally to this research.}}

\date{%
    School of Computer Science\\%
    University College Dublin\\
    \emph{firstname.lastname@ucd.ie}\\[2ex]
    \today
}




\maketitle
\begin{abstract}
 There has been some concern about the impact of predatory publishers on scientific research for some time. Recently, publishers that might previously have been considered `predatory' have established their \emph{bona fides}, at least to the extent that they are included in citation impact scores such as the field-weighted citation impact (FWCI). These are sometimes called `grey' publishers (MDPI, Frontiers, Hindawi). In this paper, we show that the citation landscape for these grey publications is significantly different from the mainstream landscape and that affording publications in these venues the same status as publications in mainstream journals may significantly distort metrics such as the FWCI. 
\end{abstract}

\section{Introduction}

The world of scientific publishing has changed significantly in recent years. Two important changes are the increase in volume and the move to Open Access \cite{piwowar2018state,shu2024oligopoly}. The latter often requires the authors to pay a publishing fee, which has introduced new commercial opportunities and attracted new players into the publishing business. Some of these new publishers have been branded as `predatory' \cite{siler2020demarcating} because they are viewed to be exploiting the Open Access model without following the best practices for academic publishing to ensure scientific quality. They often solicit and readily accept manuscripts by offering rapid review and publication, which can be problematic for maintaining the high scholarly standards needed to ensure lasting quality.

In this paper, we present an observational study of ~1.4M computer science (CS) journal papers published between 2000 and 2023. The collection is based on the CS bibliography maintained on the DBLP website (\url{dblp.org}), and the citation data was obtained from Semantic Scholar (\url{semanticscholar.org}). We find that the number of journal publications per year in CS has increased from 20,000 in 2000 to 120,000 in 2023. While some of this increase can be accounted for by the `traditional' journals from publishers such as Elsevier or IEEE, by 2023 one-third of publications now come from the new Open Access journals. This volume growth alone is concerning, but does it affect publication quality? To answer this question, we explore whether there are differences in quality between publications in new Open Access venues and those in traditional journals. Our analysis accounts for the move to \emph{hybrid} models that include OA among traditional publishers 
 in order to distinguish between these and newer \emph{OA-first} publications. 


Our main findings are:
\begin{itemize}
    \item We show a rapid increase in publication volume and an even more rapid increase in citations. 
    \item We also report a surprising increase in the average number of authors on CS papers; we later discuss possible reasons for this and whether or not it may be problematic.
    \item We show that, on average, the impact of papers in New OA journals is not equivalent to the impact of publications in traditional venues.
    \item Indeed, when traditional journals switch to a hybrid model (by accommodating OA papers) the impact of these OA papers holds up. 
    
\end{itemize}

The paper proceeds as follows. In the next section, we provide a brief history of the emergence of Open Access publications in CS. In section \ref{sec:metrics}, we review the main metrics used to quantify research impact. Section \ref{sec:data} provides details on the dataset used in our analysis. Section \ref{sec:OverallTrends} presents the overall trends evident in the dataset, and section \ref{sec:impact} discusses what can be learned about differences in citation impact from the data. The paper finishes with some recommendations for robust citation metrics in section \ref{sec:recs} and a summary and some conclusions in section \ref{sec:conc}.



\section{Open Access Publication}

Once the arrival of the internet made online publication possible, the prospect of Open Access (OA) publication became a reality \cite{bailey1991}. The principles underlying OA publication were laid out in the Budapest Open Access Initiative in 2001 \cite{bailey2017introduction}: research literature would be freely available online, either in pre-print or after peer-review; it was expected that copyright would reside with the authors.

OA would change the economics of academic publishing. It would remove the need for subscriptions or once-off payments to access research literature. It was expected that online publication would greatly reduce production costs, and these reduced costs could be borne by initiatives within research communities. Two such initiatives in Computer Science were the emergence of the Journal of AI Research (JAIR) and the Journal of Machine Learning Research (JMLR). The JAIR was launched online in 1993. Paper copies were then printed by AAAI Press but are now published by the AI Access Foundation, a nonprofit charity. In 2001, 40 editorial board members of the Machine Learning journal resigned, citing a lack of harmony between the needs of the machine learning community -- principally the need for the more efficient circulation of published papers -- and the business model of commercial publishers \cite{atkeson2001editorial}. This led directly to the creation of JMLR to better serve the needs of the research community by providing more immediate and universal access to articles.

There are at least two distinct forms of OA publishing \cite{beall2013predatory}. With \emph{Gold OA} the final published version of research articles is permanently and freely available and the authors will likely have paid an article processing charge (APC). JMLR and JAIR would be unusual in this regard, these are Gold OA journals without an APC.  With \emph{Green OA}, a version of the article (but probably not the final version) is available in a repository such as arXiv, and no APC will have been paid. 

OA benefits scientific progress and democratises research by making publications universally available. Because of this, research funding agencies increasingly require research results to be available as OA. However, despite the manifest benefits of OA, the initiative is an interesting example of the law of unintended consequences in action. Instead of readers paying for access to research papers, there has been a significant move towards researchers paying a significant and upfront charge for publication. This has created a large market for paid academic publication, with new opportunities to exploit researchers and academics who must survive in the increasingly competitive publish-or-perish academic world. Not surprisingly, this has attracted many new unscrupulous players into this market. Beall \cite{beall2013predatory} argues that the Gold OA model has a fatal flaw: it motivates publishers to publish more and more but without the guardrails of scientific quality or research merit. 

Indeed, the emergence of unethical practices in Gold OA has led to the widespread use of the terms
`predatory journal' and `predatory publishers' to describe this phenomenon \cite{grudniewicz2019predatory,oviedo2021journal}. A predatory publisher is a publisher whose main motivation is the financial gain that can be derived from a high-volume OA business model. The problem is exacerbated because the demarcation between legitimate and predatory publications is increasingly unclear. Publishers such as Hindawi, Frontiers and MDPI, which might have been branded predatory in the past, now host journals in their portfolios that are included in the main research indexing services such as PubMed and Scopus. Because of the emergence of this `grey area` between traditional and predatory journals,  some publishers are referred to as `grey' rather than `predatory'\cite{siler2020demarcating}. The landscape is further complicated because some traditional publishers have set up their own Gold OA journals to benefit from this increasingly lucrative market. IEEE Access is perhaps the most visible example of this, and for this reason, in our analysis, we distinguish between traditional IEEE journals and IEEE Access, grouping the latter with other grey publishers such as Hindawi, Frontiers, and MDPI. To further complicate things many traditional journals have adopted a hybrid model where papers are published Gold OA subject to an APC. Happily, our analysis suggests that there are no quality issues with these OA papers in hybrid journals.

\section{Publication Metrics}\label{sec:metrics}
It is worth emphasising that the best way to evaluate the quality of a research paper is to read it carefully \cite{west2010eigenfactor}. At the same time, methods for quantifying the significance or impact of a paper do have an important and established role; for example, if we wish to read up on a new research area, then the \emph{most widely cited} papers are a good place to start. The practice of ranking journals based on citation metrics is well established and relied upon by researchers, academic institutions, and research funding agencies as an important quality indicator. At the same time it must be emphasized that publication metrics are using citations as an indicator or proxy for research quality. There will be cases where good quality research is not picked up by these metrics. 
In this section, we present an overview of these metrics, focusing on metrics that might be impacted by an increase in publication volume and variability in publication quality. 

\subsection{Field Weighted Citation Impact}
The Field Weighted Citation Impact (FWCI) is a key metric in the SciVal research performance assessment service operated by Elsevier \cite{purkayastha2019comparison}. The metric is based on a weighted citation count and it is applied to a collection of papers, so it can be used to compare researchers, journals or institutions. A `field weighting' of 1 indicates the world citation average for a particular field, while a score that is less than or greater than 1 indicates a citation count that is lower than or higher than the field average, respectively. More formally, for a collection of $N$ papers, the FWCI is based on the following equation\cite{purkayastha2019comparison}:
\begin{equation}
    FWCI = \frac{1}{N}\sum_{i=1}^{N}\frac{c_i}{e_i}
\end{equation}
The $c_i$ term is the number of citations received by publication $i$ in the publication year plus the following three years, and $e_i$ is a normalising factor representing the \emph{expected} number of citations in the same period. SciVal allocates publications to categories and sub-categories, and $e_i$ is calculated based on other papers in the same sub-category. 
Importantly, FWCI does not differentiate between publishers, so MDPI or Elsevier papers within the same sub-category are treated in the same manner. 

\subsection{$h$-Index}
The $h$-index was first proposed by Hirsch \cite{hirsch2005index} in 2005 as an author-level metric to measure productivity and citation impact. It is based on a scientist's most cited papers and the number of citations that they have received. A scientist with $N_p$ papers has index $h$ if $h$ of their $N_p$ papers have at least $h$ citations each and the other ($N_p-h$) papers have $\le h$ citations each. More formally, if $f(i)$ is the number of citations for paper $i$ and if we order these from the highest to the lowest then:
\begin{equation}
    h = \max\{i \in \mathbb{N} : f(i) \ge i\}
\end{equation}

The $h$-index has several good characteristics; it is a good measure of sustained research impact, and it discourages over-publication on a particular topic or result (`salami slicing'). However, in calculating the $h$-index, all citations to a paper have the same status regardless of whether the paper is included as a passing reference to related work or whether it is cited as a fundamental result that has underpinned or influenced the citing research. In section \ref{sec:OverallTrends}, we will see that the typical bibliography in papers today is twice as long as twenty years ago, so citations must be ``easier to get'' now than they used to be. 

Another issue with the $h$-index (and with FWCI) is that it does not  account for differences in author counts. For example, with the $h$-index metric, every scholar in a multi-authored article receives full credit for its citation count, even though the contributions of all scholars are rarely equal. Accordingly, several proposed variants seek to normalise the $h$-index score based on author count \cite{bihari2023review}. Two formulae for achieving this are as follows:
\begin{equation}
    h_i = \frac{h}{Avg_A} ;\quad h_p = \frac{h}{\sqrt{Avg_A}}
\end{equation}
$h_i$ is simply the $h$-index normalised by the average author count in the papers contributing to the index. $h_p$ is the \emph{pure} $h$-index where the normalisation is the square root of the average author count -- the objective being to dampen the impact of a small number of papers with very large author counts \cite{bihari2023review}. Given that papers with very long author lists are rare in Computer Science this square root may not be necessary; we return to this issue in section \ref{sec:recs}.

\subsection{Network Centrality}\label{sec:centrality}
The problem with methods such as the $h$-index and FWCI, which depend on citation counts, is that they do not consider the quality or importance of citations. While FWCI attempts to normalise across subject areas, it treats all citations \emph{within} a subject area as equivalent. There are good reasons to treat some citations differently from others. For example, it may make sense for a citation from a high-impact paper to carry more weight than a citation from a more minor paper in the field. This issue has received much attention in research on network centrality measures. 

The practice of quantifying network centrality comes from social network analysis, where an important objective is identifying influential people (nodes) in a network \cite{landherr2010critical}. Perhaps the easiest way to do this is to count the number of links/edges pointing to a node in a network; in a social network, the edges/links refer to friendship or other social relationships between nodes (people). This is a node's \emph{in-degree centrality}; \emph{degree} refers to the edge count. In citation networks, degree centrality corresponds to a count of citations (edges) between papers (nodes) without consideration of citation significance. If nodes A and B each have five citations, they have the same influence score, as measured by degree centrality. However, if the citations for A come from nodes with high influence scores, we may wish to consider that by weighting these links more highly than citations from nodes with lower influence scores.  Eigenvector centrality accounts for this. 

Eigenvector centrality uses an adjacency network in which nodes are individuals, and the edges are links between these individuals. A citation network \emph{is} an adjacency network where the papers are nodes and the edges are links. Adjacency networks can be represented as square matrices where entry $(i,j)$ represents a link between nodes $i$ and $j$. One conceptualisation of eigenvector centrality is that it represents the probability of a random walk ending up on a particular node; intuitively, well-connected nodes should have higher probabilities. The largest eigenvector of the adjacency matrix represents these probabilities. 

Two variants of eigenvector centrality are PageRank and Katz centrality. PageRank has had a huge impact on information retrieval because of its ability to consider the importance of incoming links. It is worth noting that PageRank was inspired by ideas coming from research in citation analysis \cite{brin1998anatomy}. Katz centrality is a compromise between degree and eigenvector centrality because it includes an attenuation factor to down-weight the impact of influential nodes \cite{wkas2018axiomatization}. In our analysis in section \ref{sec:impact}, we use eigenvector centrality.

In 2008, Bergstrom \emph{et al.} introduced their Eigenfactor Metrics \cite{bergstrom2008eigenfactor} as a concrete proposal for scoring journal impact using network centrality. Their Eigenfactor Score quantifies how important a journal is in a citation networks in terms of network centrality. Because this gives larger journals more influence, they propose the Article Influence Score which normalises the Eigenfactor Score by the number of articles published in the journal. 

Another journal metric that uses network centrality is the SJR score\cite{gonzalez2010new}; it has the advantage that the SJR scores for most journals are freely available online at (\url{scimagojr.com}).

\subsection{Influential Citations}\label{sec:InfCit}
We will see in the analysis in section \ref{sec:OverallTrends} that there has been a significant increase in the reference count in papers with the move to online. So, more than ever, there is a need to identify which references are the most meaningful and which have had a significant impact on the cited work. Semantic Scholar attempts to do this by using a `Highly Influential Citations' measure. This is based on the work by Valenzuela and Etzioni
\cite{valenzuela2015identifying}, where they propose a supervised Machine Learning method for identifying citations that either use or extend the cited work -- these are labelled as influential citations -- in contrast with citations that cite related work or that are included for comparison purposes. 

In the original Valenzuela and Etzioni work found that approximately 15\% of citations were influential; the proportions we report in section \ref{sec:OverallTrends} are somewhat lower than this. Crucially, this metric helps to differentiate between the New OA and other publications.

\section{The Data}\label{sec:data}
The dataset used in this study was generated using DBLP and Semantic Scholar (April-May, 2024). An core set of papers was collected from DBLP by extracting all the journal articles listed, resulting in a collection of 1,896,440 unique articles. Thus DBLP defines the set of journal articles covered. 


These articles were used to download corresponding Semantic Scholar records to obtain their citations\footnote{Due to the nature of the Semantic Scholar API we are limited to a maximum of 1,000 citations per paper.}. The Semantic Scholar records for each citation were also downloaded, resulting in 11,723,390 seed and citing articles. For this study, we focus on the 1,389,339 articles published from 2000 up to 2023; we refer to this as the \emph{seed set}

Next, we defined a set of \emph{Traditional} publishers -- IEEE (excluding IEEE Access), Elsevier, Springer, ACM) -- and a set of \emph{New OA} publishers -- Hindawi, IEEE Access, Frontiers, MDPI -- to distinguish between publishers adopting more conventional academic publication practices versus more recent paid, OA models. We identified all articles in our dataset published in these target journals and further distinguished between OA articles from traditional publishers, which we refer to as \emph{Trad (OA)}, and non-OA articles from traditional publishers, which we refer to as \emph{Trad}. This resulted in a subset of 755,890 \emph{Trad} articles, and 379,313 \emph{Trad(OA)} articles, and 254,136 \emph{New OA} articles; these articles are associated with 11,548,876 unique citations. 

We use this 2000-2023 Trad/Trad(OA)/New OA dataset in two ways: (i) to analyse overall trends in the CS publication landscape since 2000, and (ii) to analyse the citation impact of traditional versus (new) OA publishers. Since traditional publishers have a longer history and, therefore, a greater opportunity to attract citations, for this impact analysis, we focus on a smaller subset of 850,369 articles published between 2015 and 2023;  391,602 Trad, 229205 Trad (OA), and  229,562 New OA articles.

\section{Overall CS Publication Trends}\label{sec:OverallTrends}
\begin{figure}[t]  
\includegraphics[width=\linewidth]{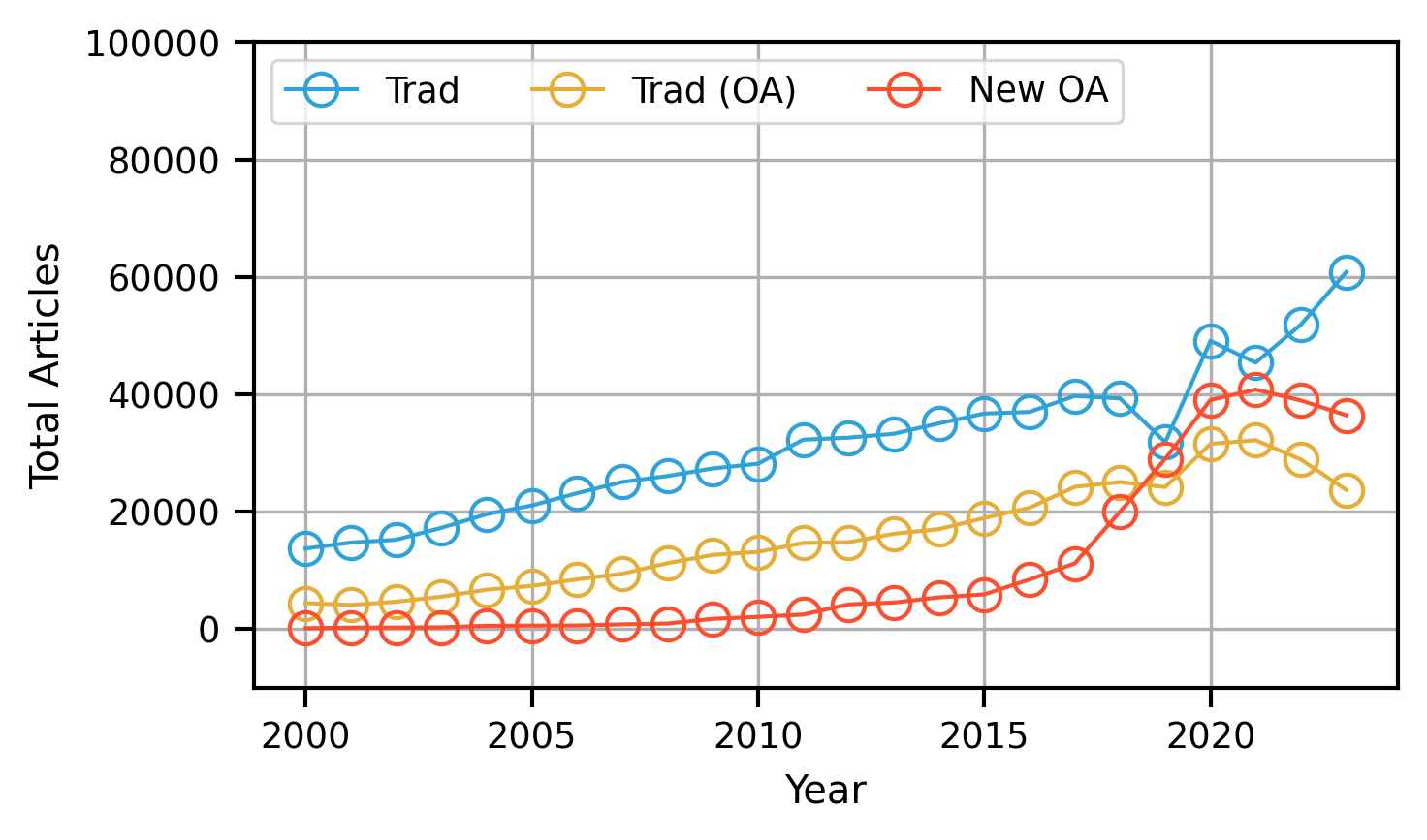} 
      \caption{The publication output of both traditional and New OA publishers has increased significantly since 2000.\label{fig:pub_count}}
\end{figure}

\begin{figure*}
    \centering
    \begin{subfigure}[t]{0.49\textwidth}
        \centering
        \includegraphics[width=\linewidth]{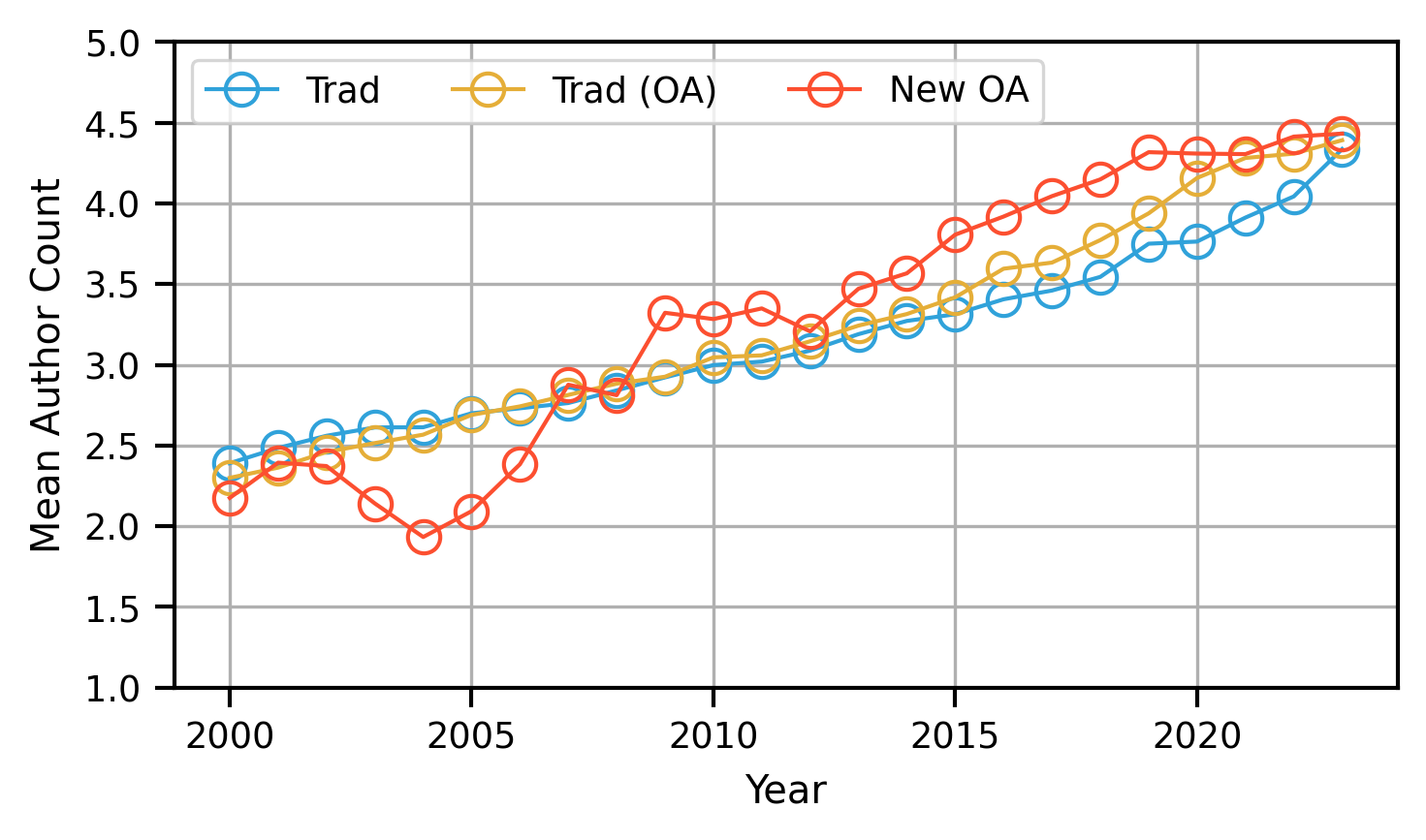} 
        \caption{Author count} \label{fig:auth_count}
    \end{subfigure}
    \begin{subfigure}[t]{0.49\textwidth}
        \centering
        \includegraphics[width=\linewidth]{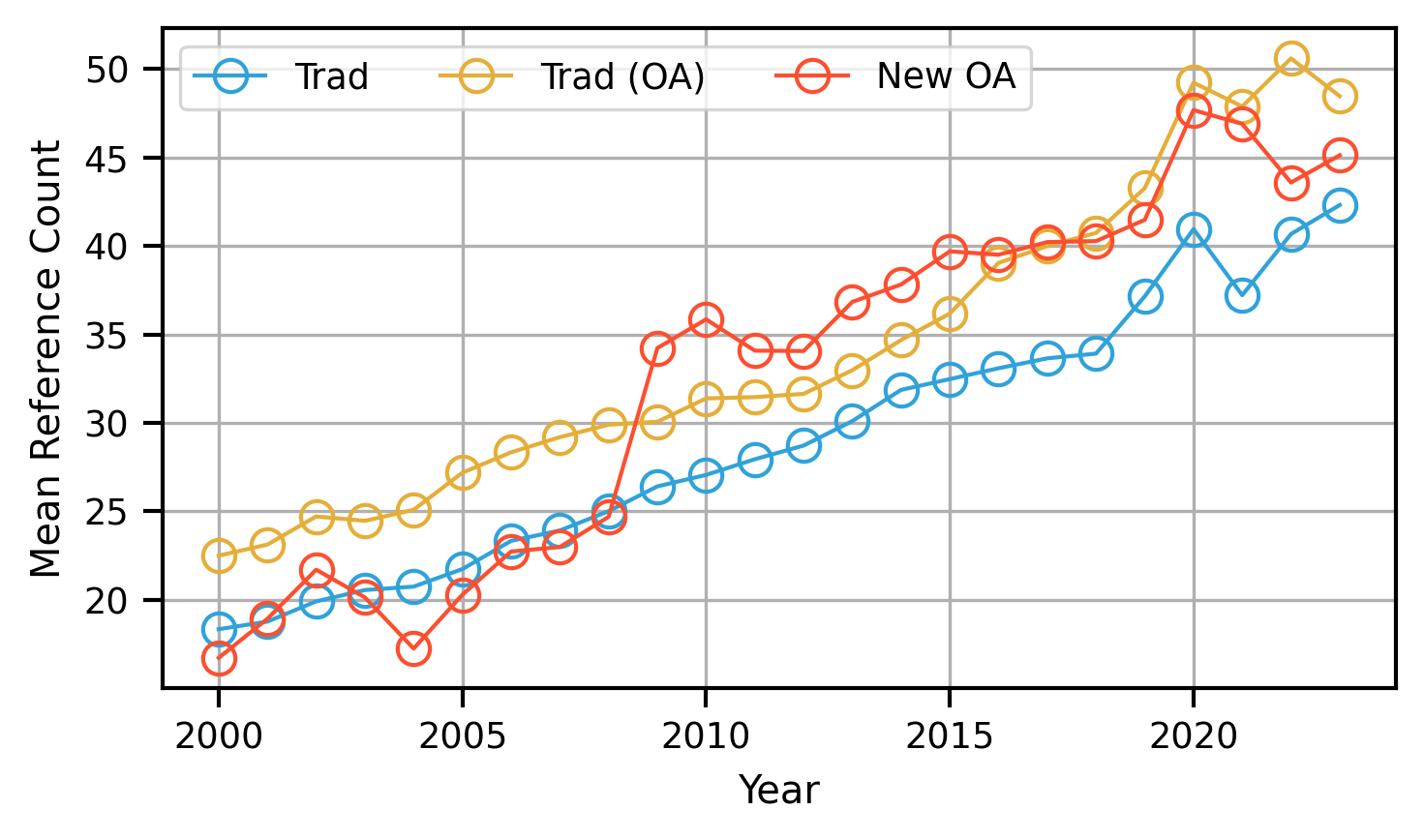} 
        \caption{Average reference count} \label{fig:ref_count}
    \end{subfigure}
    \hfill
    \begin{subfigure}[t]{0.49\textwidth}
        \centering
\includegraphics[width=\linewidth]{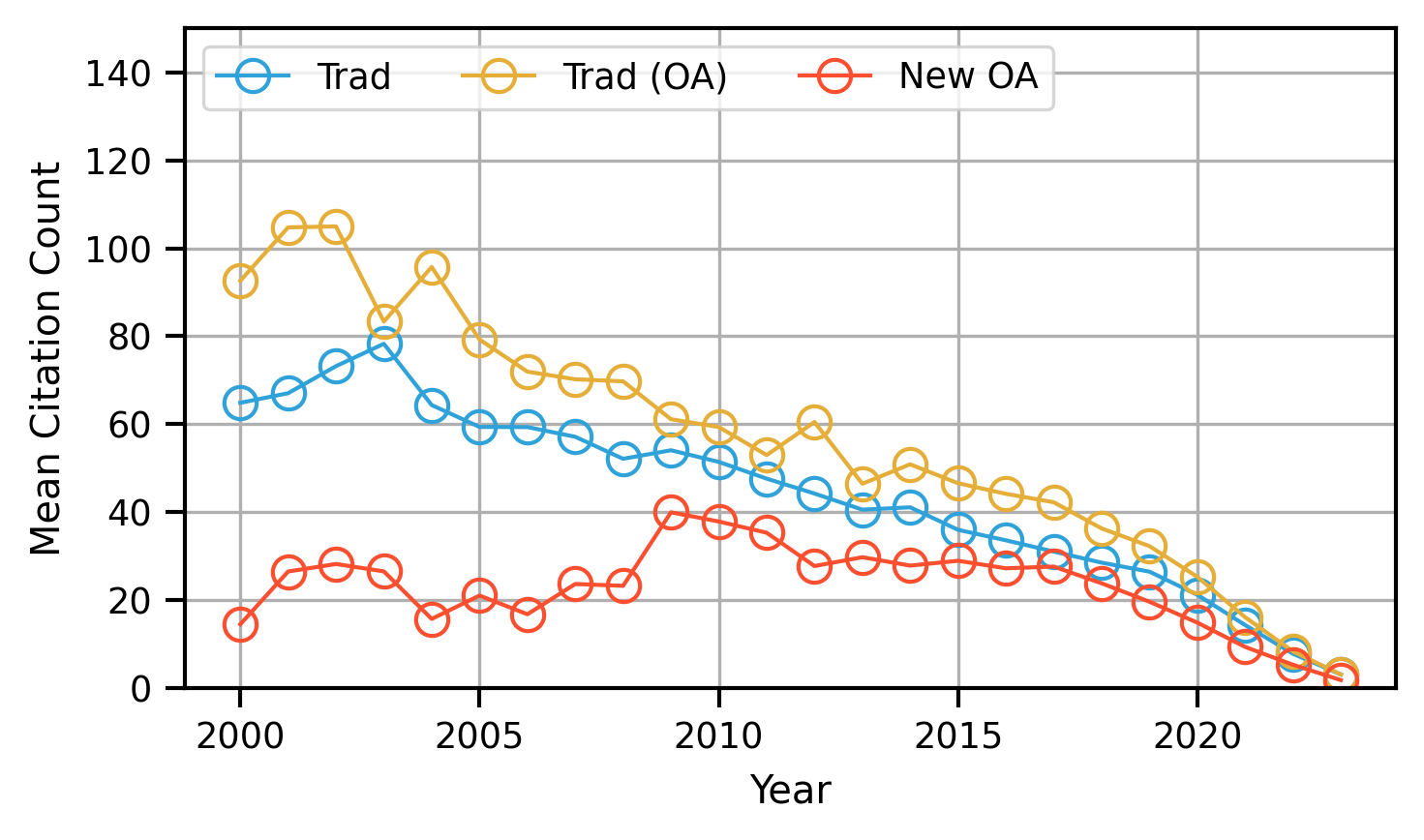} 
        \caption{Average citation count} \label{fig:cit_count}
    \end{subfigure}
    \begin{subfigure}[t]{0.49\textwidth}
        \centering
        \includegraphics[width=\linewidth]{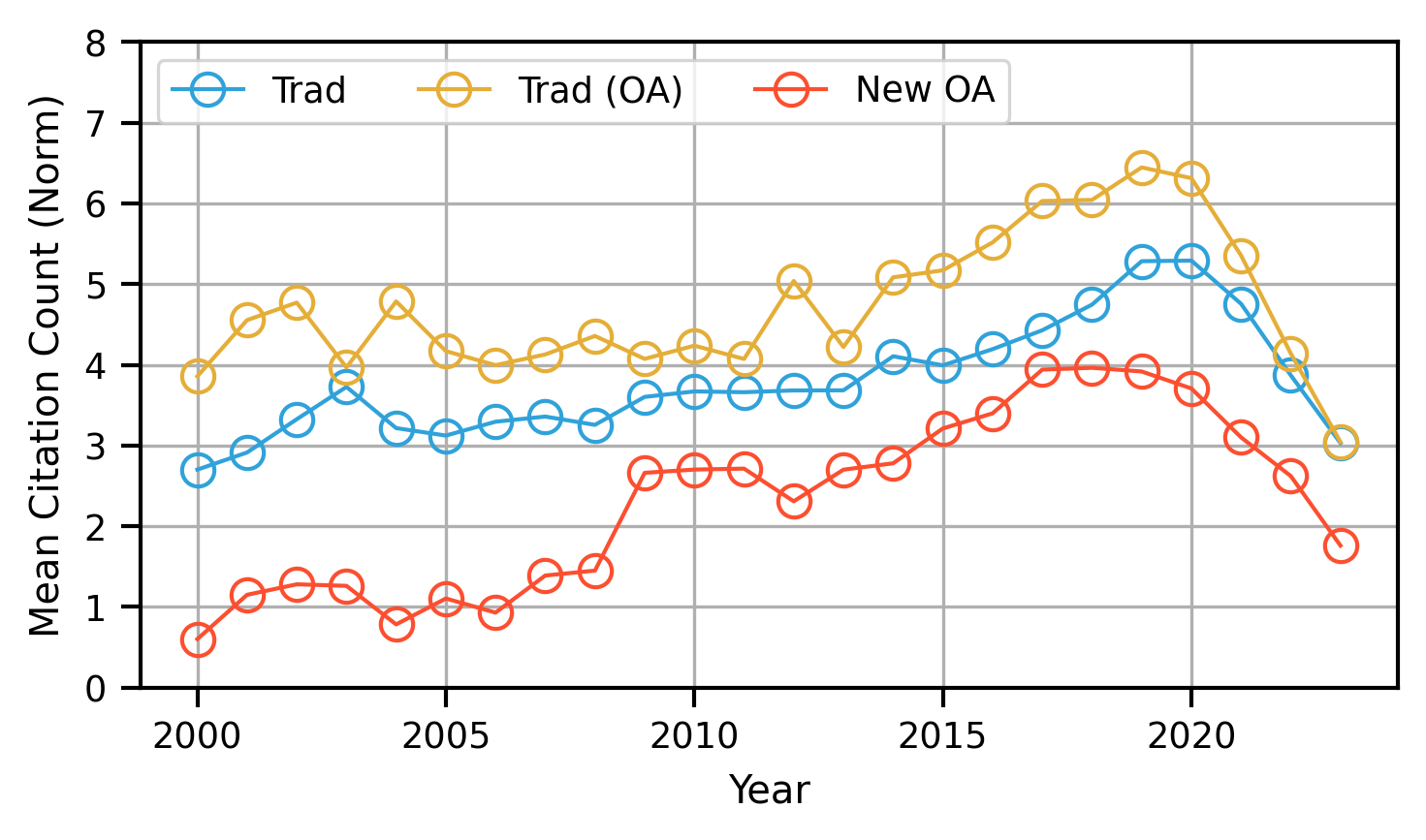} 
        \caption{Normalised average citation count} \label{fig:norm_cit_count}
    \end{subfigure}
    \begin{subfigure}[t]{0.49\textwidth}
        \centering \includegraphics[width=\linewidth]{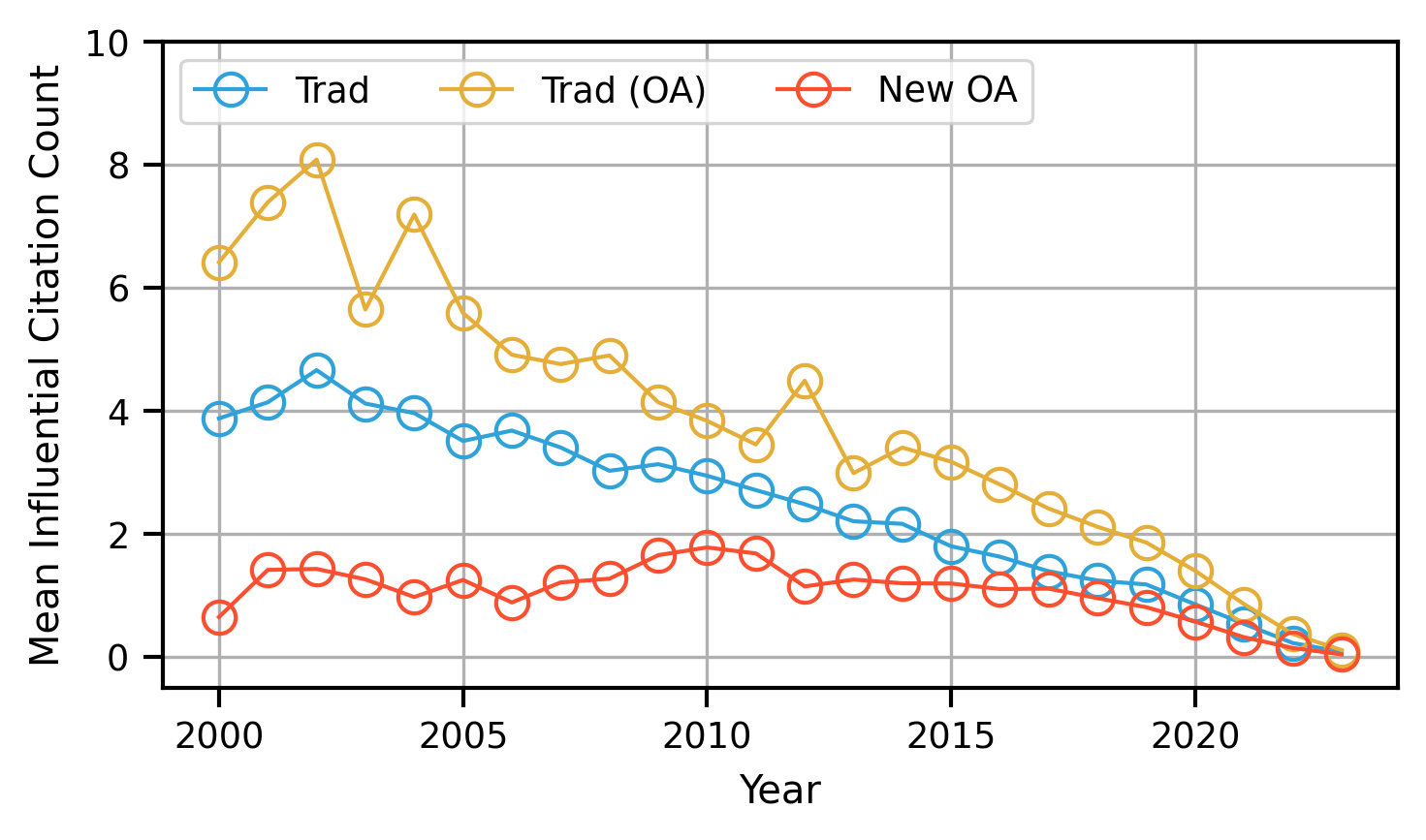} 
        \caption{Influential citation count} \label{fig:inf_cit_count}
    \end{subfigure}
    \hfill
    \begin{subfigure}[t]{0.49\textwidth}
        \centering
        \includegraphics[width=\linewidth]{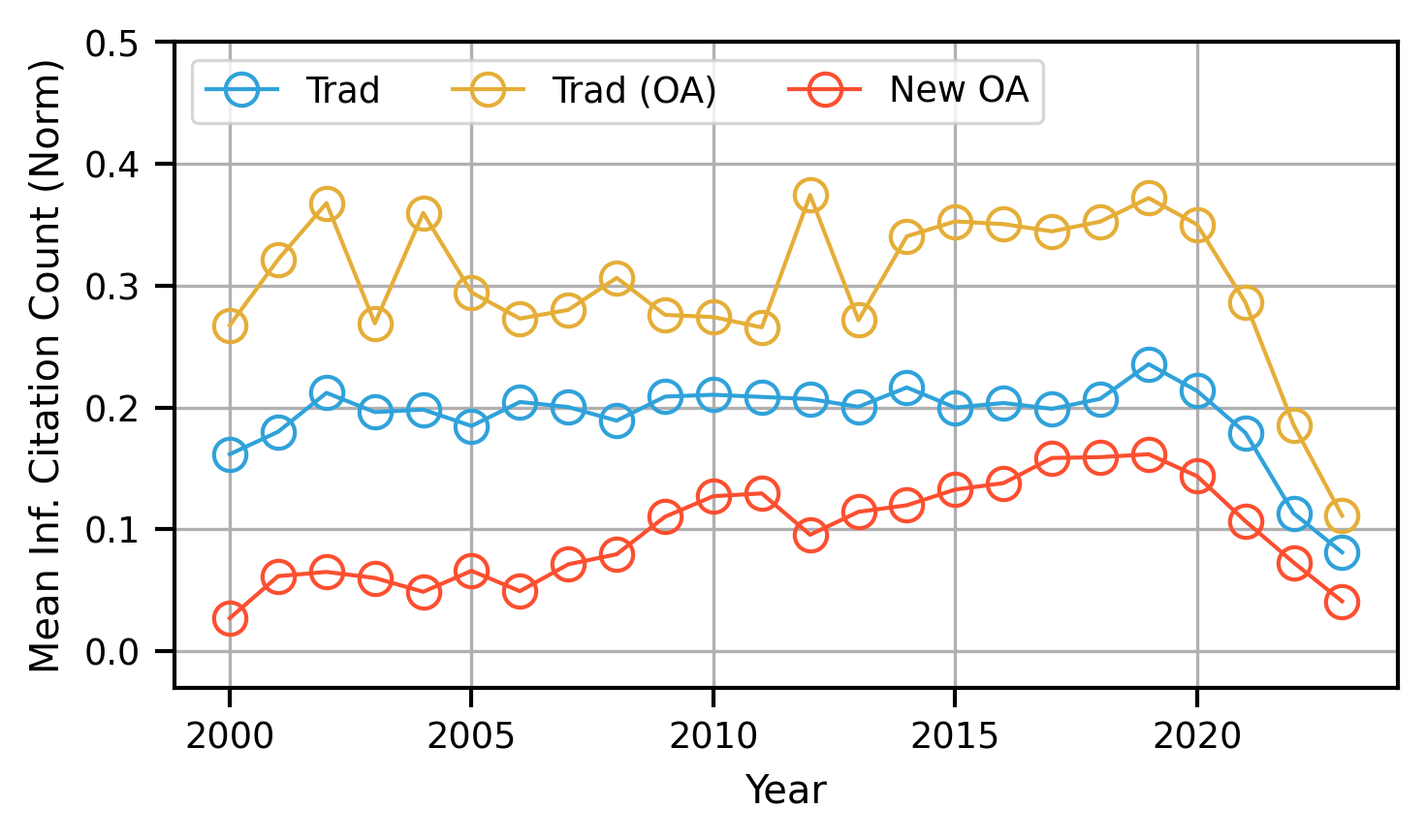} 
        \caption{Normalised influential citation count.} \label{fig:inf_cit_count_norm}
    \end{subfigure}

\caption{Overall statistics for the 2000 to 2023 paper collection ($\sim1.4$M papers).}
\label{fig:overall_stats}
\end{figure*}

This section looks at the overall trends in the CS publication landscape since 2000, comparing New OA (Hindawi, MDPI, Frontiers, IEEE Access) and Traditional (IEEE, ACM, Elesvier, Springer) publishers. Altogether this comprises $\sim$ 1.4M papers. The main trends in this dataset are shown in Figures \ref{fig:pub_count} and \ref{fig:overall_stats} and the following observations can be made.
\begin{itemize}
    \item \textbf{Paper Count:}
The first thing to note is the significant increase in publication volume (Figure \ref{fig:pub_count}). The output from the traditional publishers has grown steadily from about 20K papers in 2000 to over 80K p.a. in 2023. Significant output from New OA publishers emerged after 2008 and increased steeply in the five years up to 2020, reaching 40K papers p.a.

There is a noticeable decline in the 2019 data for Traditional publishers; it is most obvious for \emph{Trad} but also evident in \emph{Trad (OA)}. The 2020 numbers for \emph{Trad} appear to be higher than expected, which may suggest that the previous year's dip is due to a data issue with Semantic Scholar whereby some 2019 articles are incorrectly listed as 2020 articles. We do not believe that this issue has a material impact on our analysis of the results.

\item \textbf{Author Count:}
One of the more surprising trends in the dataset is shown in Figure \ref{fig:auth_count}. There has been a steady growth in the average number of authors per paper, growing from just over 2 authors per paper in 2000 to more than 4 authors per paper in 2023. This trend is evident in both the traditional and New OA collections. It is surprising because, unlike increases in references and citation counts, there is no obvious reason for it. This issue is discussed further in section \ref{sec:AuthorCount}.

\item \textbf{Reference Count:} Figure \ref{fig:ref_count} shows that the average number of references (cited papers) has been increasing steadily since 2000. This may be due to the move to online publication and the consequent relaxation of bibliographic limits in papers. However, it is noteworthy that this growth has continued since the early 2000s. While there is a sharp increase in the number of references in the New OA group from 2008 -- coinciding with the rise of New OA papers in Figure \ref{fig:pub_count} -- which probably can be explained by more relaxed bibliographic limit, the reasons for the more stable growth among traditional journals appears less clear-cut. It is noteworthy too that the mean reference count for \emph{Trad (OA)} papers exceeds that of \emph{Trad} papers, just like the \emph{New (OA)} papers.

\item \textbf{Citation Count:} The plot in Figure \ref{fig:cit_count} suggests that the two collections differ in terms of their mean citation count, but the picture is complicated by the fact that the citation count is lower for newer papers. Therefore, in Figure \ref{fig:norm_cit_count}, we present a normalised version of citation count by dividing by the number of years since publication. New OA papers have historically attracted a lower average citation count than traditional venues, and while the situation improved somewhat around 2009, the difference remains. Moroever, the \emph{Trad (OA)} papers have attracted a higher citation count than the non-OA \emph{Trad} papers, consistent with the notion that OA papers are more readily discoverable than non-OA papers.

\item \textbf{Influential Citation Count} The difference between New OA and traditional publications is more apparent when we look at influential citation count in Figures \ref{fig:inf_cit_count} and \ref{fig:inf_cit_count_norm}; once again we present a normalised measure of influential citation count (\ref{fig:inf_cit_count_norm}) by using the number of years since publication as the normalisation factor. Papers published in traditional venues attract significantly more influential citations than those published in OA venues. And, once again, \emph{Trad (OA)} outperform \emph{Trad} papers by this measure.
\end{itemize}

In summary, there are several interesting and concerning things to note. Publication volume has increased dramatically. The number of citations generated is increasing even more dramatically. \emph{New OA} papers attract fewer citations than those in traditional venues, but \emph{Trad OA} papers attract more citations that \emph{New OA} and \emph{Trad} papers. Moroever, that the citations that the \emph{New OA} papers do attract are much less likely to be influential, while the \emph{Trad (OA)} citations are much more likely to be influential. This suggests an impact disparity that is due to the publisher type (grey versus traditional publishers) rather than the publishing model per se. Finally, the average number of authors per paper has almost doubled since 2000.

\subsection{Increase in Publication Volume}
\begin{figure}[ht]
\includegraphics[width=\linewidth]{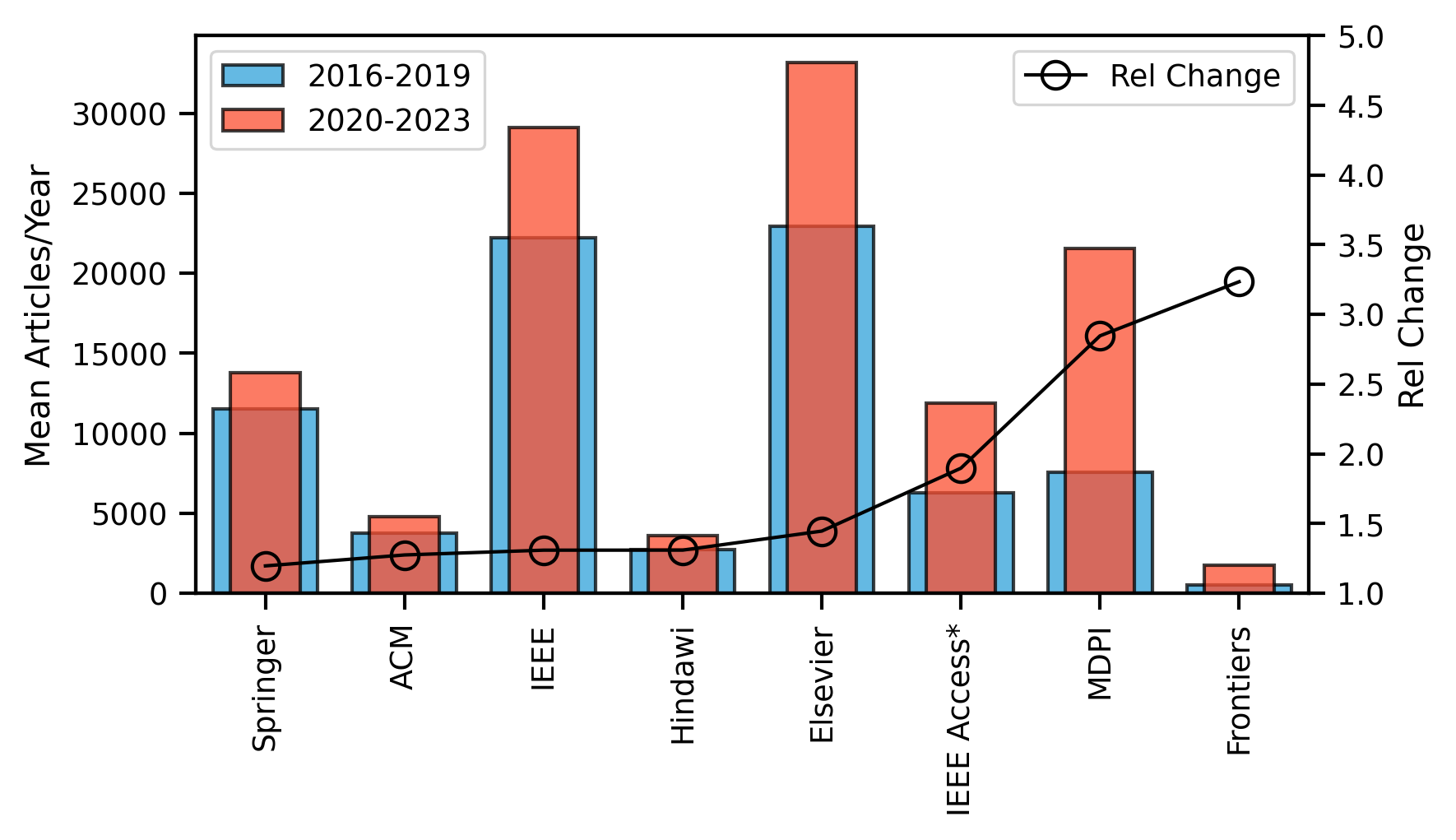}
\caption{Growth in number of publications by publisher.}
\label{fig:PaperCountYearsAlt}
\end{figure}

In Figure \ref{fig:pub_count} we see that the number of publications has increased dramatically since 2000. To look at this in more detail, we look at the relative growth for different publishers since New OA became a significant factor from 2015 onward. In Figure \ref{fig:PaperCountYearsAlt} we show changes in publication volume by publisher for the two four-year periods before and after 2000. 

The largest overall publishers are Elsevier and IEEE, in both periods, but the publishers with  largest relative growth are Frontiers, MDPI and IEEE Access; since Frontiers is coming from a low base, its overall annual output remains small despite its 3.5$\times$ growth rate. It is worth noting that the current annual volume of journal papers from MDPI since 2020 is greater than that from Springer and the ACM combined, while IEEE Access now approaches the annual output of Springer and exceeds that of the ACM. 

These volume changes have the potential to significantly reshape the CS landscape because the growth rate of the New OA publishers is more than twice that of traditional publishers. MDPI is now the number three producer of journal articles in CS, and even if it does not manage to sustain its current growth rate, even a reduced growth rate will likely be sufficient for it to overtake IEEE and perhaps Elsevier to become the largest producer of CS journal articles over the next four years.

\subsection{The Increase in Author Count}\label{sec:AuthorCount}
Whereas the increase in publication output and bibliography sizes is generally known, the increase in author count is somewhat surprising. Wu \emph{et al.} have shown that there has been a shift towards larger research teams in recent years \cite{wu2019large}. However, they also show that new ideas are likely to come from small teams, with the work from larger teams being more incremental. So, a move to larger author counts does not necessarily benefit science. 

We expect the average number of authors on papers to vary widely across disciplines, but there is no obvious reason for this authorship growth in Computer Science. It may be due to a move towards multi-disciplinary research or larger research groups, which may be characteristic of modern CS research. It may be due to increased collaboration between academic and industry partners. 

As data-driven methods have significantly impacted the nature of science across many disciplines, it may be natural to find computer science authors as additional members of publication teams, with CS papers including scientists from other domains or from industry. This will continue as artificial intelligence and machine learning will likely play an increasingly important role in future research across all disciplines. 

At the same time, if a `valid' number of authors on a paper in 2000 is two, is an average of four in 2023 equally valid?
Almost all publishers and research institutions have guidelines on co-authorship \cite{osborne2019authorship}. As a general principle, co-authors should make a significant intellectual contribution to the work. If the co-authorship count increases, is this principle being observed? If not, why might the primary authors be willing to share authorship with others if their intellectual contribution does not merit it? One possible explanation is that it reflects a desire to influence the future citation count of published papers \cite{wu2019large}. If self-citation increases citations, then the more co-authors, the more co-author-level self-citations. 

This issue of apportioning credit to authors on papers with multiple authors has been identified in research on citation analysis \cite{bihari2023review}. Does an author on a paper with six authors deserve the same credit as an author on a similar paper with two authors? As previously mentioned, metrics such as FWCI or the $h$-index do not consider this. This is a greater issue now than it was previously because there is more variation in author count. Indeed, modified versions of the $h$-index have been proposed that seek to normalise based on author count to address this issue. 

\begin{figure}[ht]
\includegraphics[width=8cm]{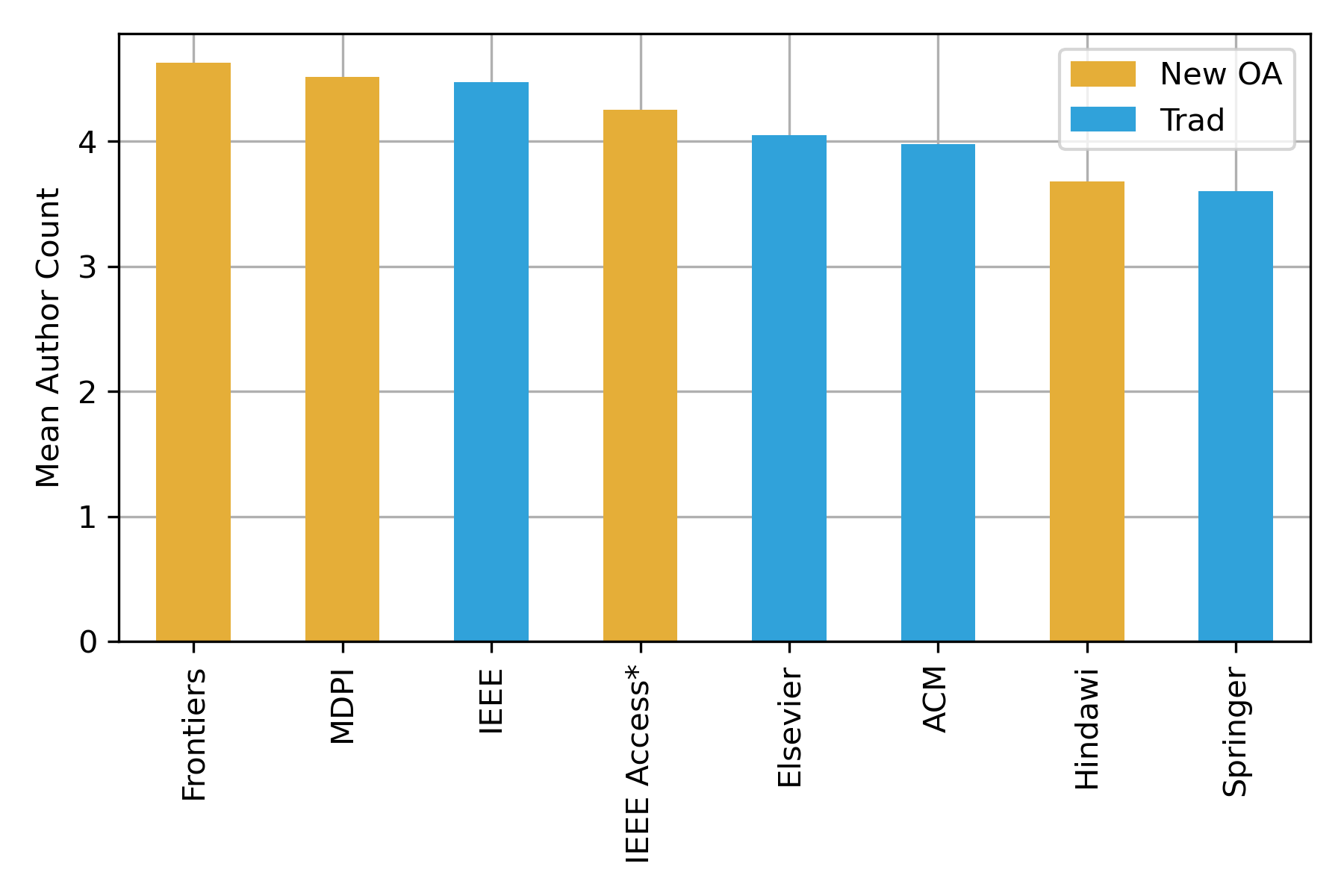}
\caption{Mean Author Counts by publisher for papers published between 2020 and 2023.}
\label{fig:MeanAuthCountByPub}
\end{figure}

To understand this change in authorship practice further, we have examined the data in more detail. Figure \ref{fig:MeanAuthCountByPub} presents the mean author count by publisher for the four years from 2020 to 2023. While three of the top four publishers are New OA, it is clear that the variation is not huge. 

However, when we look at the data at the journal level, a clearer picture starts to emerge - see Table \ref{tab:avg-aut-count}. The first thing to notice is that most of the top 10 journals, in terms of average author count for 2020 to 2023, are interdisciplinary journals; larger interdisciplinary teams can account for more authors. When we look back at the journals featured in the top 10 for 2015 to 2018, we see journals such as the Internet of Things that are effectively single-discipline. So, to some extent, the increase in author count in CS can be attributed to a move to more interdisciplinary research, which is a good thing. At the same time, it is worth noting that for the five journals appearing in the top 10 for both periods, the average author count has increased in all cases.

Thus, we can conclude that there is a common trend toward larger authorship counts across all publishers. While some of this can be explained by `legitimate' reasons, such as larger multi-disciplinary research teams, this is not always true. Therefore, it will be important to consider the impact of increasing authorship counts in the future, and since FWCI and $h$-index cannot do this, it will be important to consider variations in these metrics that can account for such trends; this issue is discussed in section \ref{sec:recs}.

\begin{table*}[]
\caption{The top 10 journals in terms of average author count for 2020-2023; the top 10 for 2015-2018 are shown for comparison.}
\centering
\label{tab:avg-aut-count}
\begin{tabular}{l|ll|ll}
\textbf{}                                         & \multicolumn{2}{l|}{\textbf{2020-2023}}               & \multicolumn{2}{l}{\textbf{2015-2018}}                \\ \hline
\textbf{Journal}                                  & \multicolumn{1}{l|}{\textbf{Mean}} & \textbf{Rank} & \multicolumn{1}{l|}{\textbf{Mean}} & \textbf{Rank} \\ \hline
Computing and Software for Big Science            & \multicolumn{1}{l|}{10.0}             & 1             & \multicolumn{1}{l|}{6.7}              & 4             \\
Genomics, Proteomics \& Bioinformatics            & \multicolumn{1}{l|}{9.3}              & 2             & \multicolumn{1}{l|}{}                 &               \\
Medical Image Analysis                            & \multicolumn{1}{l|}{8.0}              & 3             & \multicolumn{1}{l|}{6.0}              & 7             \\
IBM Journal of Research \& Development           & \multicolumn{1}{l|}{7.9}              & 4             & \multicolumn{1}{l|}{7.4}              & 2             \\
NeuroImage                                        & \multicolumn{1}{l|}{6.8}              & 5             & \multicolumn{1}{l|}{6.0}              & 8             \\
IEEE Transactions on Medical Imaging              & \multicolumn{1}{l|}{6.7}              & 6             & \multicolumn{1}{l|}{}                 &               \\
IEEE Journal of Solid-State Circuits              & \multicolumn{1}{l|}{6.5}              & 7             & \multicolumn{1}{l|}{}                 &               \\
Int. Jnl. of Comp. Assisted Radiology \& Surgery & \multicolumn{1}{l|}{6.3}              & 8             & \multicolumn{1}{l|}{6.0}              & 9             \\
Journal of Imaging Informatics in Medicine        & \multicolumn{1}{l|}{6.3}              & 9             & \multicolumn{1}{l|}{}                 &               \\
Frontiers in Digital Health                       & \multicolumn{1}{l|}{6.3}              & 10            & \multicolumn{1}{l|}{}                 &               \\
Machine Translation                               & \multicolumn{1}{l|}{}                 &               & \multicolumn{1}{l|}{8.5}              & 1             \\
Astronomy \& Computing                           & \multicolumn{1}{l|}{}                 &               & \multicolumn{1}{l|}{7.0}              & 3             \\
Internet of Things                                & \multicolumn{1}{l|}{}                 &               & \multicolumn{1}{l|}{6.5}              & 5             \\
ACM Transactions on Computer Systems              & \multicolumn{1}{l|}{}                 &               & \multicolumn{1}{l|}{6.4}              & 6             \\
Frontiers in Neuroinformatics                     & \multicolumn{1}{l|}{}                 &               & \multicolumn{1}{l|}{6.0}              & 10           
\end{tabular}
\end{table*}

\section{Impact Differences} \label{sec:impact}

An important question in this paper relates to the changes in the CS publication landscape, particularly the increase in volume. Specifically, we are interested in whether publications and citations can be counted the same. We compare New OA and traditional publishers, on a publisher by publisher basis, under two criteria: citation network centrality (introduced in Section \ref{sec:centrality}) and influential citations (Section \ref{sec:InfCit}). The analysis is based on publications since 2015 because it is clear from Figure \ref{fig:pub_count} that New OA publishers were not particularly active before 2015. 

\subsection{Network Centrality Analysis}
As per Section \ref{sec:centrality}, network centrality allows us to identify influential nodes in a citation network with papers as nodes and the citations as edges/links. The in-links to a node are citations, and the out-links are references. The result is a network with 850k nodes (papers) and 17M citations (edges), stored as an 850k $\times$ 850k matrix. 

The network can be aggregated so that the nodes are journals, with edges weighted by the number of citations from one journal to another. This network can be further aggregated so that the nodes are publishers; this publisher network has only eight nodes.

\begin{figure}[ht]
\includegraphics[width=9cm]{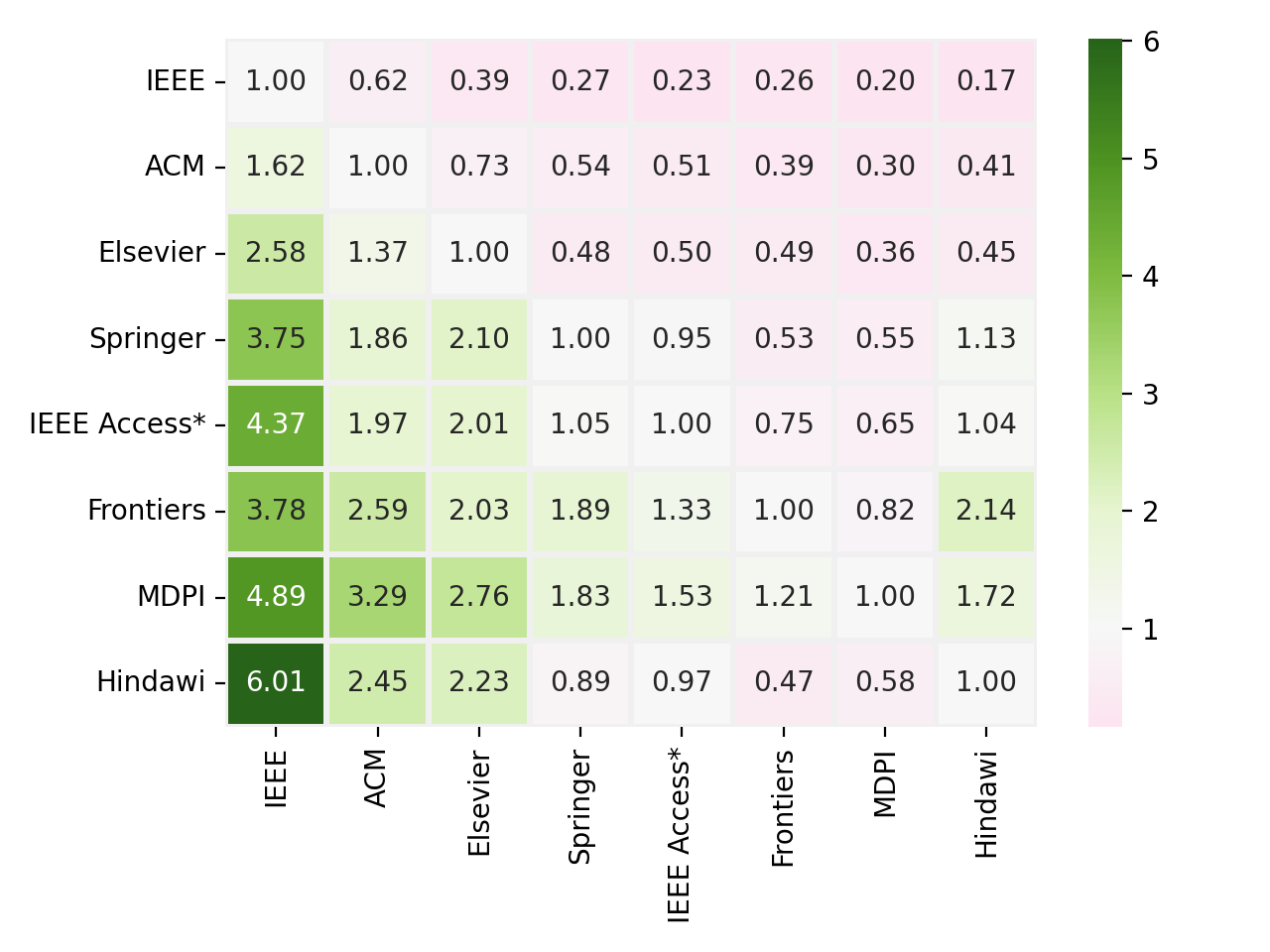}
\caption{The ratio of in-links (citations) to out-links (references) between publishers (since 2015), e.g. an IEEE paper receives 6 times the number of citations from Hindawi papers and vice versa. The heatmap colors are centred on 1.}
\label{fig:PubHeatmap}
\end{figure}

Estimating centrality in this publisher network is complicated because the volume of publications differs across publishers. In the period in question (2015 onwards), the dataset contains ~250k Elsevier papers. For comparison, there are 120k MDPI papers and 37k from ACM. So the high-volume publishers will dominate centrality measures based on raw citation counts. Therefore, we consider the ratio of in-links to out-links for our analysis. This is shown in the heatmap in Figure \ref{fig:PubHeatmap}. We are interested in citations \emph{into} a node so the numbers indicate the ratio of citations from the column publisher to the row publisher. Larger numbers (green) in the column for a publisher indicate a higher influence. For example, the bottom left cell indicates that on average an IEEE paper receives 6.01 as many citations from a Hindawi paper as a Hindawi paper received from an IEEE paper. Likewise the disparity between an ACM and MDPI is 3.29; ACM papers receive 3.29 as many links from MDPI as MDPI papers receive from ACM. And, in general, the larger values shown in the lower left quadrant indicate that traditional publishers receive many more citations from New OA publishers than New OA publishers receive from traditional publishers.

\begin{figure}[h]
\includegraphics[width=8cm]{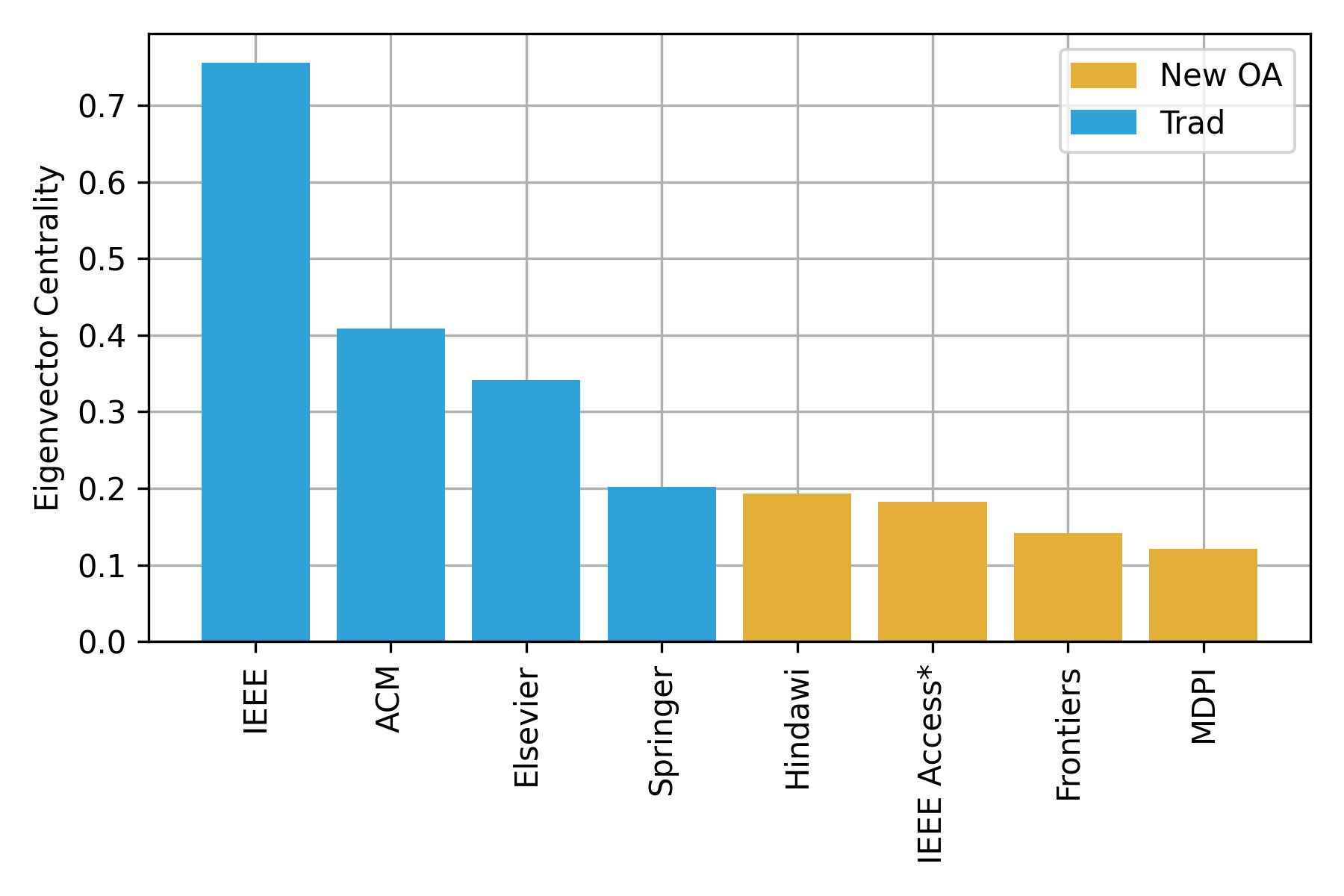}
\caption{The eigenvector centrality of publishers in the graph shown in the heatmap in Figure\ref{fig:PubHeatmap}.}
\label{fig:PubCentrality}
\end{figure}

Given that Figure \ref{fig:PubHeatmap} represents a weighted network of the normalised citation flows between publishers, we can quantify the centrality of nodes in this network using eigenvector centrality. These scores are shown in Figure \ref{fig:PubCentrality}. 
While eigenvector centrality scores are presented here, the Katz centrality scores are similar. This effectively sums up the picture in Figure \ref{fig:PubHeatmap} and shows a clear difference between the traditional publishers (IEEE, ACM, Elsevier, Springer), which have eigenvector centrality scores between 0.2 and over 0.7, compared to the much lower centrality scores (<0.2) of the New OA publishers. Despite the publication volume of MDPI and IEEE Access they do not have a significant impact. It also highlights that IEEE and IEEE Access are very different. 
 
\subsection{Citations and Influential Citations}
Online publishing has made citations easier to come by; there are more papers and more references per paper. This is why Semantic Scholar differentiates between influential and regular citations. Based on the work of \cite{valenzuela2015identifying}, citations are marked as influential if they use or extend the cited work. The alternative is an `incidental' citation, for instance, included in a list of related work.  

\begin{figure}[ht]
\includegraphics[width=\linewidth]{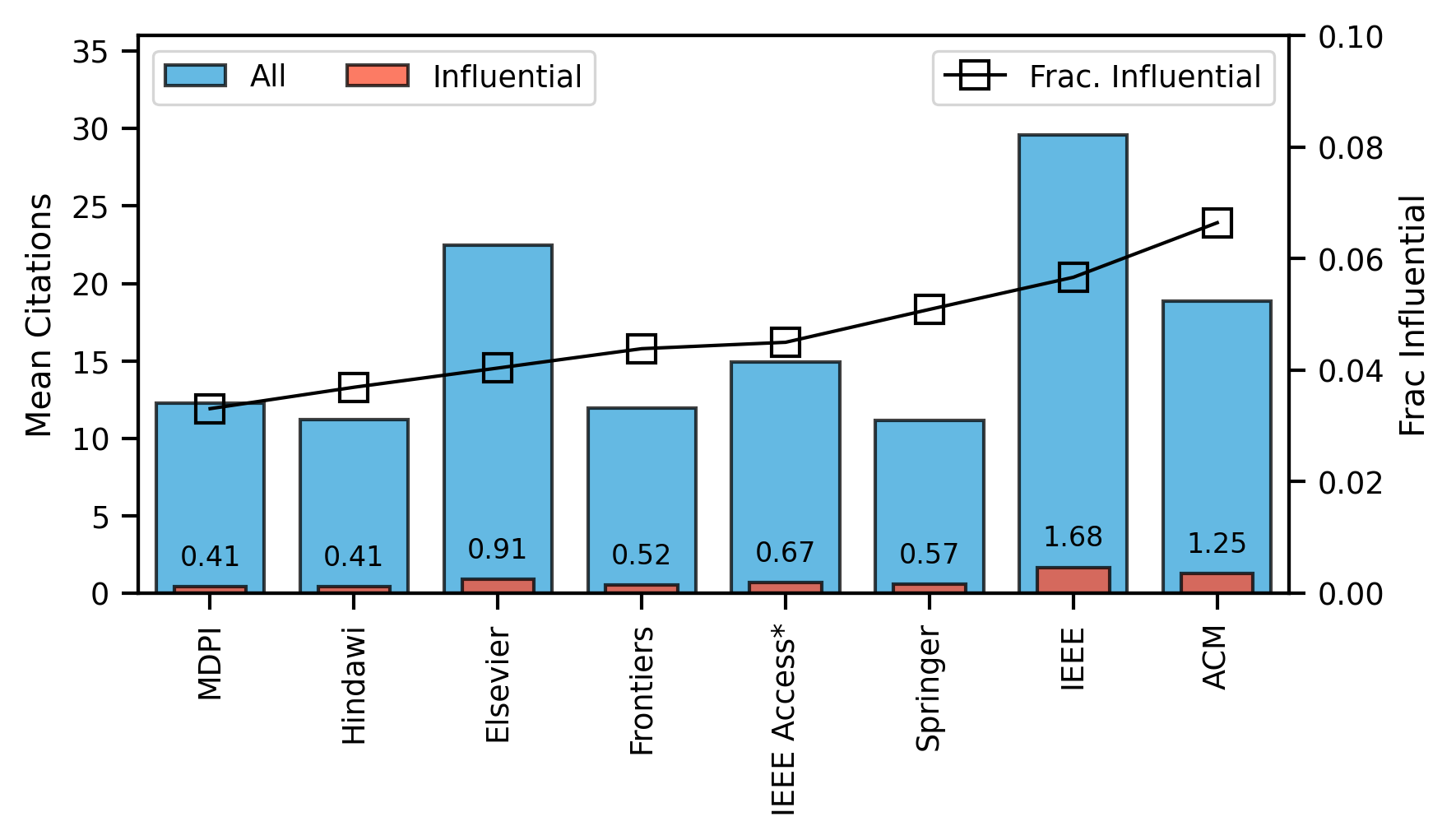}
\caption{Status of published papers: mean citation and influential citation count per paper by publisher. The black line shows the proportion of influential citations.}
\label{fig:InfCites}
\end{figure}

In our post-2015 dataset, there are 850,368 papers
with 17,121,561 citations. Of these, 824,002 are marked as influential citations, that is 4.8\% of the total. This is lower than the 
14.6\% figure in the original work by Valenzuela and Etzioni. There are several possible explanations for this difference. It might be due to the original work covering a different domain (computational linguistics). Or it could be because it is based on pre-2015 papers with more influential citations. Or their training data may be biased towards papers with at least some influential citations; as an aside, in our dataset, if we only consider papers with at least 1 influential citation, then approximately 12\% of citations are influential, which more closely matches the Valenzuela and Etzioni number. Figure \ref{fig:InfCites} shows how this fraction varies across publishers. At the high end, ACM papers are above 6\%, and MDPI is at the bottom with 3.5\%. Three of the bottom four publishers are New OA publishers. 

\begin{figure}[ht]
\includegraphics[width=\linewidth]{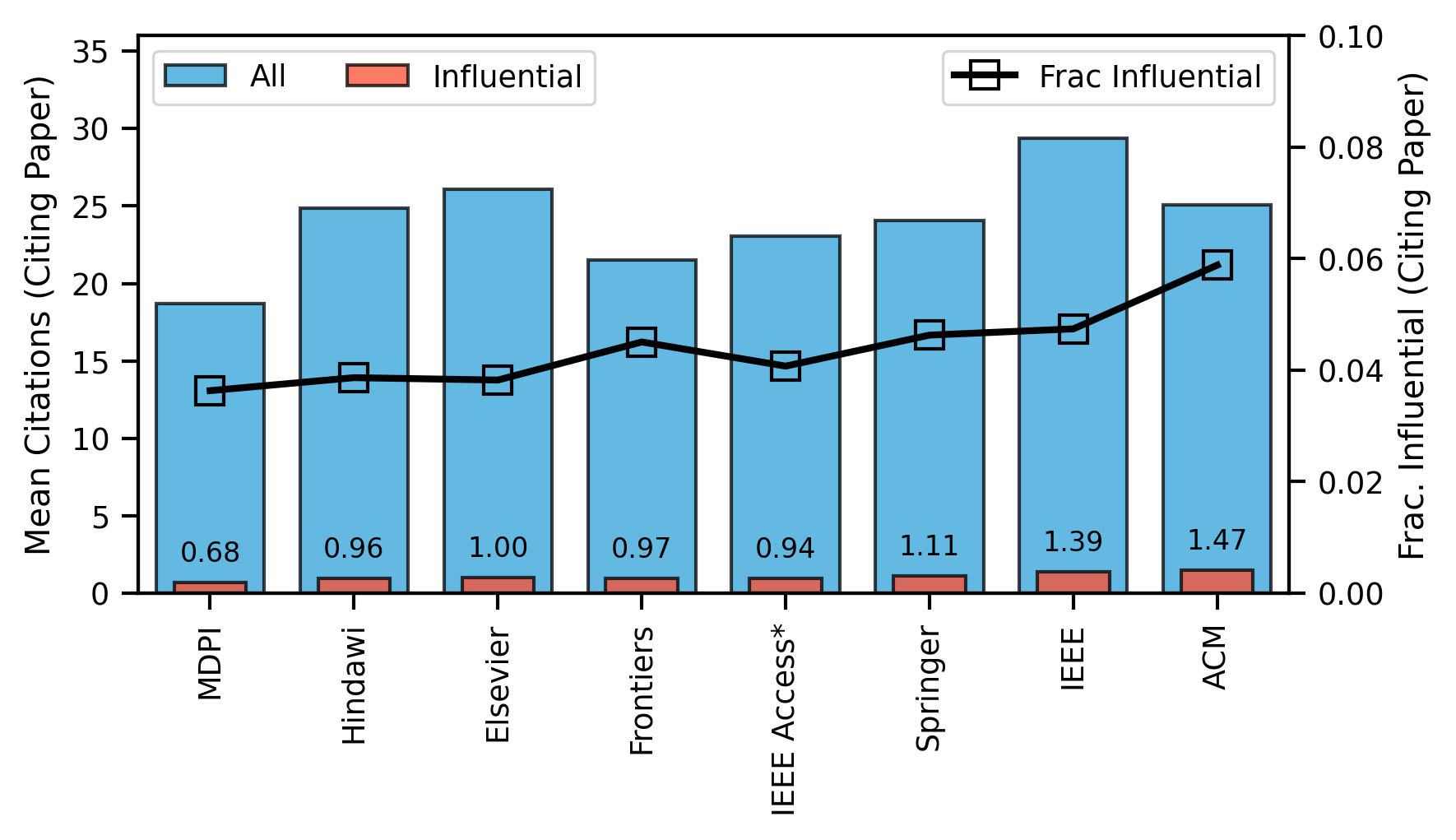}
\caption{Status of citing papers: mean citation and influential citation count for citing papers by publisher.The black line shows the proportion of influential citations.}
\label{fig:InfCitesCitePapers}
\end{figure}

In Figure \ref{fig:InfCitesCitePapers}
we look at the status of the citing papers (not the papers receiving the cites) and, in particular, \emph{their} fraction of influential cites. In this case, the variation is less, but the overall trend remains the same. At the top, papers that cite ACM papers tend to have higher fractions of influential citations compared with papers that cite MDPI or Hindawi papers, for example. Once again, three of the bottom four publishers by this measure are New OA publishers. 

Looking at overall citation counts in  Figure \ref{fig:InfCitesCitePapers}, we see that the IEEE papers come out on top with an average of almost 30 citations per paper. Elsevier and the ACM are placed second and third, with IEEE Access in fourth position. Indeed, IEEE Access fares better than the other New OA publishers in this citation analysis regarding overall citation count and the proportion of influential cites. 

\subsubsection{Statistical Significance}
Given the volume of data behind the results presented in Figures \ref{fig:InfCites} and \ref{fig:InfCitesCitePapers} it would be expected that the differences we see are statistically significant. We have tested the statistical significance of the differences in mean citation and mean influential citation counts for the eight publishers. First, we run one-way ANOVA tests to check if any pair of means is different. For both ANOVAs, the $p$-values are effectively zero suggesting that at least some of the publishers have different population means. This allows us to run a Tukey  honestly significant difference (HSD) test to make pairwise comparisons between publishers. The detailed results are presented in the Appendix. Some headline results are as follows:
\begin{itemize}
    \item The mean citation count for Elsevier papers is almost twice that of MDPI (effect size); this is statistically significant (\
    $p$=0.0).
    \item The mean citation count for traditional IEEE papers is almost twice that of IEEE Access; this is statistically significant ($p$=0.0).
    \item The mean influential citation count for ACM papers is three times that of MDPI ($p$=0.0).
\end{itemize}

\section{More Robust Metrics}\label{sec:recs}

It is clear from our analysis that the validity of established publication metrics such as the $h$-index and FWCI is open to question because they are insensitive to author count or the source of a citation. We have seen that the average author count has doubled since 2000; there is now a strong case for taking author count into consideration. The eigenvector centrality and influential citation analyses in section \ref{sec:impact} show that it is not reasonable to treat all citations the same as happens with the $h$-index and FWCI. 

Based on the review of citation metrics presented in section \ref{sec:metrics} we already have mechanisms for addressing these issues. There is well-established research on how to normalise metrics such as the $h$-index to adjust for author count \cite{bihari2023review}. A recent paper in PLOS ONE\cite{koltun2021h} demonstrates that an author's $h$-index is no longer a good indicator of research quality as evidenced by  awards that indicate recognition by the scientific community. Instead a fractional allocation of $h$-index based on author count is a better indicator. 

If citations are to be used as an indicator of quality then it should be clear from the analysis here that not all citations have the same status. Metrics based on network centrality such as the Eigenfactor Score and the Author Influence Score \cite{bergstrom2008eigenfactor} do scale citations based on influence and provide a better indicator of the impact of a research article. As an example, when the SJR score mentioned in section \ref{sec:centrality} is used to compare MDPI Sensors with the Communications of the ACM it reveals an interesting picture: according to Scopus, both journals have similar cites per paper, yet the SJR scores are respectively 0.79 and 2.96. Thus, this SJR score captures the differences illustrated in Figures \ref{fig:PubHeatmap} and \ref{fig:PubCentrality}.

It is worth noting that the Clarivate Journal Citation Reports do include the Eigenfactor Score and Author Influence scores that help to quantify the status of citations. However, at present these are provided as secondary resources; their wider use would address the issues highlighted here.

\section{Conclusions}\label{sec:conc}
The CS publication landscape has changed radically in recent years. Journal output has increased from 20,000 to 120,000 papers per year from 2000 to 2023. This is partly due to a push to improve publication availability via OA publication models, and today, a significant proportion of all papers are from New OA publishers.  However, there are concerns about a New OA downside, because the economics of OA publishing may incentivise quantity over quality. In this paper, we find evidence to support this:

\begin{itemize}
    \item New OA publishers have experienced much higher growth rates than their traditional counterparts.
    \item On average New OA papers in CS receive fewer citations than they produce; in network centrality terms New OA publishers have low status. 
    \item New OA papers receive less influential citations than traditional journals. The citations they receive are from papers that have fewer influential citations than average.
    \item OA papers from traditional publishers do not suffer in this way, in fact there is evidence that they attract more citations than their non-OA counterparts.
\end{itemize}

This suggests that any evaluation of publication quality needs to account for these changes in the publication landscape.  In particular, rather than continuing to rely on metrics such as FWCI and $h$-index, which treat all citations as equal, it may be appropriate to use methods (perhaps based on network centrality) that do not treat all citations the same \cite{portenoy2017leveraging}. 

Furthermore, while the trend towards greater authorship numbers may be justified in the case of interdisciplinary research, it cannot be so easily dismissed for more conventional research papers, especially if it is motivated by the desire to increase personal publication and citation counts. At least for non-interdisciplinary research, this trend should not be encouraged, and one way to do this is to change how we measure publication impact, perhaps by using metrics that distribute credit based on the number and positioning of authors on a paper.

In summary, given the increased volume and visibility of New OA papers, it may be important to educate researchers, especially young, early-career researchers—those most acutely exposed to the pressure of publish-or-perish—about the significant changes in the publication landscape. Moreover, the evidence reported here justifies changes in how publication quality is assessed to properly account for these landscape changes by considering authorship and impact more carefully.

\section*{Acknowledgements}
Supported by Science Foundation Ireland through the Insight Centre for Data Analytics (12/RC/2289\_P2).

\bibliographystyle{abbrv}
\bibliography{CS_Pubs}

\appendix

\begin{table*}[]
\caption{The results of a Tukey HSD test on pairwise comparisons between the mean citation counts of publishers.}
\begin{tabular}{lllll}
\toprule
 & p-value & Statistic & Lower CI & Upper CI \\
Comparison &  &  &  &  \\
\midrule
ACM-Elsevier & 0.000 & -3.586 & -5.068 & -2.105 \\
ACM-Frontiers & 0.000 & 6.929 & 3.839 & 10.018 \\
ACM-Hindawi & 0.000 & 7.692 & 5.571 & 9.813 \\
ACM-IEEE & 0.000 & -10.681 & -12.173 & -9.189 \\
ACM-IEEE Access* & 0.000 & 3.970 & 2.272 & 5.669 \\
ACM-MDPI & 0.000 & 6.616 & 5.035 & 8.198 \\
ACM-Springer & 0.000 & 7.731 & 6.134 & 9.328 \\
Elsevier-ACM & 0.000 & 3.586 & 2.105 & 5.068 \\
Elsevier-Frontiers & 0.000 & 10.515 & 7.700 & 13.330 \\
Elsevier-Hindawi & 0.000 & 11.278 & 9.582 & 12.975 \\
Elsevier-IEEE & 0.000 & -7.094 & -7.872 & -6.317 \\
Elsevier-IEEE Access* & 0.000 & 7.557 & 6.433 & 8.680 \\
Elsevier-MDPI & 0.000 & 10.203 & 9.265 & 11.141 \\
Elsevier-Springer & 0.000 & 11.317 & 10.354 & 12.281 \\
Frontiers-ACM & 0.000 & -6.929 & -10.018 & -3.839 \\
Frontiers-Elsevier & 0.000 & -10.515 & -13.330 & -7.700 \\
Frontiers-Hindawi & 0.996 & 0.763 & -2.435 & 3.962 \\
Frontiers-IEEE & 0.000 & -17.609 & -20.430 & -14.789 \\
Frontiers-IEEE Access* & 0.047 & -2.958 & -5.893 & -0.024 \\
Frontiers-MDPI & 1.000 & -0.312 & -3.181 & 2.557 \\
Frontiers-Springer & 0.990 & 0.802 & -2.075 & 3.679 \\
Hindawi-ACM & 0.000 & -7.692 & -9.813 & -5.571 \\
Hindawi-Elsevier & 0.000 & -11.278 & -12.975 & -9.582 \\
Hindawi-Frontiers & 0.996 & -0.763 & -3.962 & 2.435 \\
Hindawi-IEEE & 0.000 & -18.373 & -20.079 & -16.667 \\
Hindawi-IEEE Access* & 0.000 & -3.722 & -5.611 & -1.833 \\
Hindawi-MDPI & 0.602 & -1.076 & -2.861 & 0.709 \\
Hindawi-Springer & 1.000 & 0.039 & -1.760 & 1.837 \\
IEEE-ACM & 0.000 & 10.681 & 9.189 & 12.173 \\
IEEE-Elsevier & 0.000 & 7.094 & 6.317 & 7.872 \\
IEEE-Frontiers & 0.000 & 17.609 & 14.789 & 20.430 \\
IEEE-Hindawi & 0.000 & 18.373 & 16.667 & 20.079 \\
IEEE-IEEE Access* & 0.000 & 14.651 & 13.513 & 15.789 \\
IEEE-MDPI & 0.000 & 17.297 & 16.342 & 18.252 \\
IEEE-Springer & 0.000 & 18.411 & 17.431 & 19.392 \\
IEEE Access*-ACM & 0.000 & -3.970 & -5.669 & -2.272 \\
IEEE Access*-Elsevier & 0.000 & -7.557 & -8.680 & -6.433 \\
IEEE Access*-Frontiers & 0.047 & 2.958 & 0.024 & 5.893 \\
IEEE Access*-Hindawi & 0.000 & 3.722 & 1.833 & 5.611 \\
IEEE Access*-IEEE & 0.000 & -14.651 & -15.789 & -13.513 \\
IEEE Access*-MDPI & 0.000 & 2.646 & 1.393 & 3.899 \\
IEEE Access*-Springer & 0.000 & 3.761 & 2.488 & 5.033 \\
MDPI-ACM & 0.000 & -6.616 & -8.198 & -5.035 \\
MDPI-Elsevier & 0.000 & -10.203 & -11.141 & -9.265 \\
MDPI-Frontiers & 1.000 & 0.312 & -2.557 & 3.181 \\
MDPI-Hindawi & 0.602 & 1.076 & -0.709 & 2.861 \\
MDPI-IEEE & 0.000 & -17.297 & -18.252 & -16.342 \\
MDPI-IEEE Access* & 0.000 & -2.646 & -3.899 & -1.393 \\
MDPI-Springer & 0.049 & 1.114 & 0.003 & 2.226 \\
Springer-ACM & 0.000 & -7.731 & -9.328 & -6.134 \\
Springer-Elsevier & 0.000 & -11.317 & -12.281 & -10.354 \\
Springer-Frontiers & 0.990 & -0.802 & -3.679 & 2.075 \\
Springer-Hindawi & 1.000 & -0.039 & -1.837 & 1.760 \\
Springer-IEEE & 0.000 & -18.411 & -19.392 & -17.431 \\
Springer-IEEE Access* & 0.000 & -3.761 & -5.033 & -2.488 \\
Springer-MDPI & 0.049 & -1.114 & -2.226 & -0.003 \\
\bottomrule
\end{tabular}
\end{table*}

\begin{table*}
\caption{The results of a Tukey HSD test on pairwise comparisons between the mean influential citation counts of publishers.}
\begin{tabular}{lllll}
\toprule
 & p-value & Statistic & Lower CI & Upper CI \\
Comparison &  &  &  &  \\
\midrule
ACM-Elsevier & 0.000 & 0.347 & 0.159 & 0.536 \\
ACM-Frontiers & 0.000 & 0.730 & 0.337 & 1.123 \\
ACM-Hindawi & 0.000 & 0.841 & 0.571 & 1.111 \\
ACM-IEEE & 0.000 & -0.421 & -0.610 & -0.231 \\
ACM-IEEE Access* & 0.000 & 0.584 & 0.367 & 0.800 \\
ACM-MDPI & 0.000 & 0.849 & 0.647 & 1.050 \\
ACM-Springer & 0.000 & 0.687 & 0.484 & 0.890 \\
Elsevier-ACM & 0.000 & -0.347 & -0.536 & -0.159 \\
Elsevier-Frontiers & 0.026 & 0.383 & 0.025 & 0.741 \\
Elsevier-Hindawi & 0.000 & 0.494 & 0.278 & 0.710 \\
Elsevier-IEEE & 0.000 & -0.768 & -0.867 & -0.669 \\
Elsevier-IEEE Access* & 0.000 & 0.236 & 0.093 & 0.379 \\
Elsevier-MDPI & 0.000 & 0.501 & 0.382 & 0.621 \\
Elsevier-Springer & 0.000 & 0.340 & 0.217 & 0.462 \\
Frontiers-ACM & 0.000 & -0.730 & -1.123 & -0.337 \\
Frontiers-Elsevier & 0.026 & -0.383 & -0.741 & -0.025 \\
Frontiers-Hindawi & 0.992 & 0.111 & -0.296 & 0.518 \\
Frontiers-IEEE & 0.000 & -1.151 & -1.510 & -0.792 \\
Frontiers-IEEE Access* & 0.934 & -0.147 & -0.520 & 0.226 \\
Frontiers-MDPI & 0.977 & 0.118 & -0.247 & 0.483 \\
Frontiers-Springer & 1.000 & -0.043 & -0.409 & 0.323 \\
Hindawi-ACM & 0.000 & -0.841 & -1.111 & -0.571 \\
Hindawi-Elsevier & 0.000 & -0.494 & -0.710 & -0.278 \\
Hindawi-Frontiers & 0.992 & -0.111 & -0.518 & 0.296 \\
Hindawi-IEEE & 0.000 & -1.262 & -1.479 & -1.045 \\
Hindawi-IEEE Access* & 0.025 & -0.258 & -0.498 & -0.017 \\
Hindawi-MDPI & 1.000 & 0.007 & -0.220 & 0.234 \\
Hindawi-Springer & 0.452 & -0.154 & -0.383 & 0.075 \\
IEEE-ACM & 0.000 & 0.421 & 0.231 & 0.610 \\
IEEE-Elsevier & 0.000 & 0.768 & 0.669 & 0.867 \\
IEEE-Frontiers & 0.000 & 1.151 & 0.792 & 1.510 \\
IEEE-Hindawi & 0.000 & 1.262 & 1.045 & 1.479 \\
IEEE-IEEE Access* & 0.000 & 1.004 & 0.859 & 1.149 \\
IEEE-MDPI & 0.000 & 1.269 & 1.148 & 1.391 \\
IEEE-Springer & 0.000 & 1.107 & 0.983 & 1.232 \\
IEEE Access*-ACM & 0.000 & -0.584 & -0.800 & -0.367 \\
IEEE Access*-Elsevier & 0.000 & -0.236 & -0.379 & -0.093 \\
IEEE Access*-Frontiers & 0.934 & 0.147 & -0.226 & 0.520 \\
IEEE Access*-Hindawi & 0.025 & 0.258 & 0.017 & 0.498 \\
IEEE Access*-IEEE & 0.000 & -1.004 & -1.149 & -0.859 \\
IEEE Access*-MDPI & 0.000 & 0.265 & 0.106 & 0.424 \\
IEEE Access*-Springer & 0.525 & 0.103 & -0.058 & 0.265 \\
MDPI-ACM & 0.000 & -0.849 & -1.050 & -0.647 \\
MDPI-Elsevier & 0.000 & -0.501 & -0.621 & -0.382 \\
MDPI-Frontiers & 0.977 & -0.118 & -0.483 & 0.247 \\
MDPI-Hindawi & 1.000 & -0.007 & -0.234 & 0.220 \\
MDPI-IEEE & 0.000 & -1.269 & -1.391 & -1.148 \\
MDPI-IEEE Access* & 0.000 & -0.265 & -0.424 & -0.106 \\
MDPI-Springer & 0.013 & -0.162 & -0.303 & -0.020 \\
Springer-ACM & 0.000 & -0.687 & -0.890 & -0.484 \\
Springer-Elsevier & 0.000 & -0.340 & -0.462 & -0.217 \\
Springer-Frontiers & 1.000 & 0.043 & -0.323 & 0.409 \\
Springer-Hindawi & 0.452 & 0.154 & -0.075 & 0.383 \\
Springer-IEEE & 0.000 & -1.107 & -1.232 & -0.983 \\
Springer-IEEE Access* & 0.525 & -0.103 & -0.265 & 0.058 \\
Springer-MDPI & 0.013 & 0.162 & 0.020 & 0.303 \\
\bottomrule
\end{tabular}
\end{table*}

\end{document}